\documentclass[final]{svjour3}
\usepackage{graphicx}
\usepackage{rotating}
\usepackage{amssymb}
\usepackage{mathptmx}
\usepackage[numbers]{natbib}
\usepackage{color}
\usepackage{float}
\makeatletter
\journalname{Journal of Low Temperature Physics}

\bibpunct{[}{]}{,}{n}{}{,}

\begin{document}
\newcommand{\hdblarrow}{H\makebox[0.9ex][l]{$\downdownarrows$}-}
\title{The Simons Observatory: Magnetic Shielding Measurements for the Universal Multiplexing Module}

\author{Zachary B. Huber$^1$ 
    \and Yaqiong Li$^{1,2}$
    \and Eve M. Vavagiakis$^1$ 
    \and Steve K. Choi$^{1,3}$
    \and Jake Connors$^{4,5}$
    \and Nicholas F. Cothard$^6$ 
    \and Cody J. Duell$^1$ 
    \and Nicholas Galitzki$^7$ 
    \and Erin Healy$^8$ 
    \and Johannes Hubmayr$^4$ 
    \and Bradley R. Johnson$^9$ 
    \and Ben Keller$^1$ 
    \and Heather McCarrick$^8$ 
    \and Michael D. Niemack$^{1,2,3}$
    \and Yuhan Wang$^8$ 
    \and Zhilei Xu$^{10}$ 
    \and Kaiwen Zheng$^8$}

\institute{1. Department of Physics, Cornell University, Ithaca, NY 14853, USA\\
2. Kavli Institute at Cornell for Nanoscale Science, Cornell University, Ithaca, NY 14853, USA \\
3. Department of Astronomy, Cornell University, Ithaca, NY 14853, USA \\
4. Quantum Sensors Group, NIST, 325 Broadway, Boulder, CO 80305, USA \\
5. Department of Physics, University of Colorado Boulder, Boulder, Colorado 80309, USA\\
6. NASA Goddard Space Flight Center, Greenbelt, MD 20771, USA \\
7. University of California San Diego, Department of Physics, La Jolla, CA 92093 USA \\
8. Joseph Henry Laboratories of Physics, Jadwin Hall, Princeton University, Princeton, NJ 08544, USA \\
9. University of Virginia, Department of Astronomy, Charlottesville, VA 22904, USA \\
10. MIT Kavli Institute, Massachusetts Institute of Technology, 77 Massachusetts Avenue, Cambridge, MA 02139, USA \\
\email{zbh5@cornell.edu}}

\maketitle

\begin{abstract}

The Simons Observatory (SO) includes four telescopes that will measure the temperature and polarization of the cosmic microwave background using over 60,000 highly sensitive transition-edge bolometers (TES). These multichroic TES bolometers are read out by a microwave RF SQUID multiplexing system with a multiplexing factor of 910. Given that both TESes and SQUIDs are susceptible to magnetic field pickup and that it is hard to predict how they will respond to such fields, it is important to characterize the magnetic response of these systems empirically. This information can then be used to limit spurious signals by informing magnetic shielding designs for the detectors and readout.
This paper focuses on measurements of magnetic pickup with different magnetic shielding configurations for the SO universal multiplexing module (UMM), which contains the SQUIDs, associated resonators, and TES bias circuit. The magnetic pickup of a prototype UMM was tested under three shielding configurations: no shielding (copper packaging), aluminum packaging for the UMM, and a tin/lead-plated shield surrounding the entire dilution refrigerator 100 mK cold stage. We also present measurements of the pickup in the UMM when aluminum feedhorns are installed. The measurements show that the aluminum packaging outperforms the copper packaging by a shielding factor of 8-10, and adding the tin/lead-plated 1K shield further increases the relative shielding factor in the aluminum configuration by 1-2 orders of magnitude. The addition of feedhorns provides a factor of 30 improvement when the tin/lead shield is not installed and a factor of 5 improvement when it is.


\end{abstract}

\section{Introduction}

Next-generation cosmic microwave background (CMB) experiments like the Simons Observatory (SO) \cite{so_science} will utilize advances in multiplexing readout technology and high throughput optical designs to field at least four times more detectors than the previous generation of telescopes like the Atacama Cosmology Telescope (ACT) \cite{act} and the South Pole Telescope (SPT) \cite{spt}. SO will initially field over 60,000 transition-edge sensor (TES) bolometers across its 6-meter Large Aperture Telescope (LAT) and three half-meter Small Aperture Telescopes (SATs). These detectors will be read out by microwave multiplexing ($\mu$mux) circuits \cite{bdober21} that utilize resonators tuned to different frequencies and SQUIDs to read out approximately 900 detectors per RF line. Both the detectors and the readout circuit are sensitive to magnetic fields in ways that are difficult to predict or simulate. Since magnetic pickup from the movement of the telescope through the magnetic field of Earth, cryogenic half-wave plates used in the SATs, and other sources inside the telescope can produce spurious signals, it is crucial to characterize empirically the effectiveness of the magnetic shielding of the deployment-grade readout packaging to ensure that magnetic field pickup is being minimized.

Simons Observatory is packaging all of the cold readout technology into a common universal microwave-multiplexing module (UMM) that can be used with all detector arrays regardless of frequency across all of the SO telescopes \cite{mccarrick21}. Each UMM is composed of $\mu$mux chips containing the resonator-coupled SQUIDs, a silicon wafer that both connects the $\mu$mux chips in series and contains the TES bias circuitry, and an enclosing copper frame. An outer clamp surrounds the UMM on all sides except the bottom, where the readout module will be coupled to the detector array. The design for the outer clamp aims to achieve good thermal and magnetic shielding properties by gold plating a superconducting aluminum clamp. For deployment on the sky, a TES detector wafer and an aluminum feedhorn array are coupled to the UMM to make a universal focal-plane module (UFM).  

In section two, we describe the experimental setup. In section three, we present measured results for a prototype UMM in three configurations: a copper clamp, an aluminum clamp, and an aluminum clamp with a tin/lead-plated 1K shell surrounding the UMM. We also present the magnetic sensitivity of the SQUIDs in the UMM when the feedhorns have been added to make a UFM. The datasets analyzed during the current study are available from the corresponding author on reasonable request.

\vspace{-0.2in}

\section{Methods}

\vspace{-0.1in}

\begin{figure}[htbp]
\vspace{-0.2in}
\begin{center}
\includegraphics[width=0.95\linewidth, keepaspectratio]{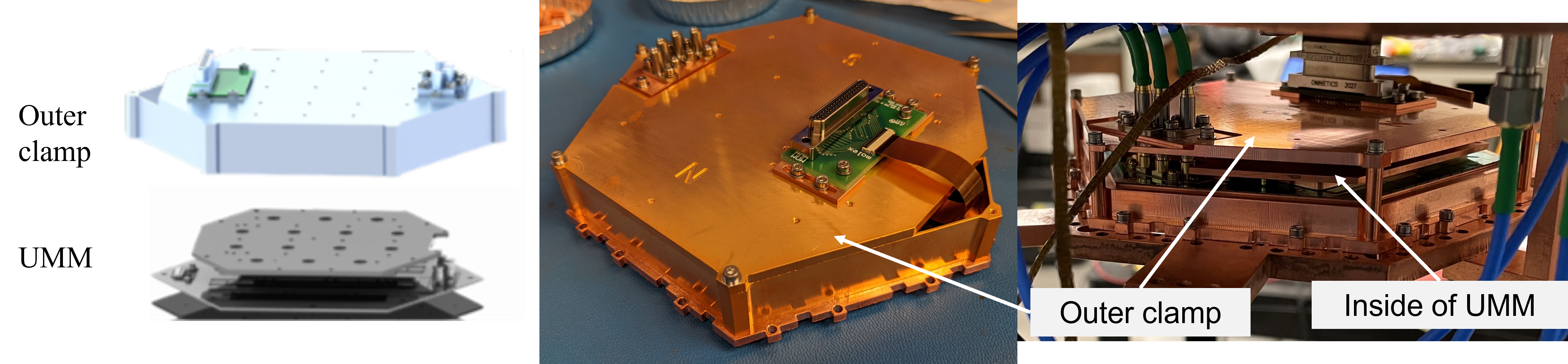}
\caption{(Color figure online.) {\it Left} - Exploded view of UMM from \cite{mccarrick21} showing the outer clamp and inner components. For these tests, only the outer clamp changes between different shielding configurations.
The two packaging configurations used for shielding measurements: {\it Center} - the gold-plated aluminum clamp; {\it Right} - the copper clamp.
\vspace{-0.2in}
\label{fig:UMMconfigs}}
\end{center}
\end{figure}

For testing the magnetic sensitivity of the SO UMM, we worked with two configurations of packaging on a prototype UMM: one with an aluminum outer clamp incorporating both lid and walls and one with a copper outer clamp using a lid only that was made as simple as possible to minimize any metallic components that could provide magnetic shielding (see Fig.~\ref{fig:UMMconfigs}). Neither UMM configuration is completely enclosed by the clamp as both clamps have a hole to allow a flexible cable to pass through, and there is no superconducting shield on the ``bottom'' (i.e. sky-side) of the UMM. When the UMM is converted to a UFM, the detector wafer stack and aluminum feedhorns are coupled at this sky-side face.

After installing the clamp, the UMM was mounted to the 100 mK stage of a Bluefors dilution refrigerator (DR). During the cooldown, the fridge was surrounded by an external high-permeability mu-metal shield to prevent magnetic flux from becoming trapped in the SQUIDs as they transition to the superconducting state. When the UMMs are deployed on SO telescopes they will be inside of a high-permeability A4K magnetic shield, 
as described in \cite{zhu21} and \cite{ali20}.

\begin{figure}[htbp]
\vspace{-0.2in}
\begin{center}
\includegraphics[width=0.95\linewidth, keepaspectratio]{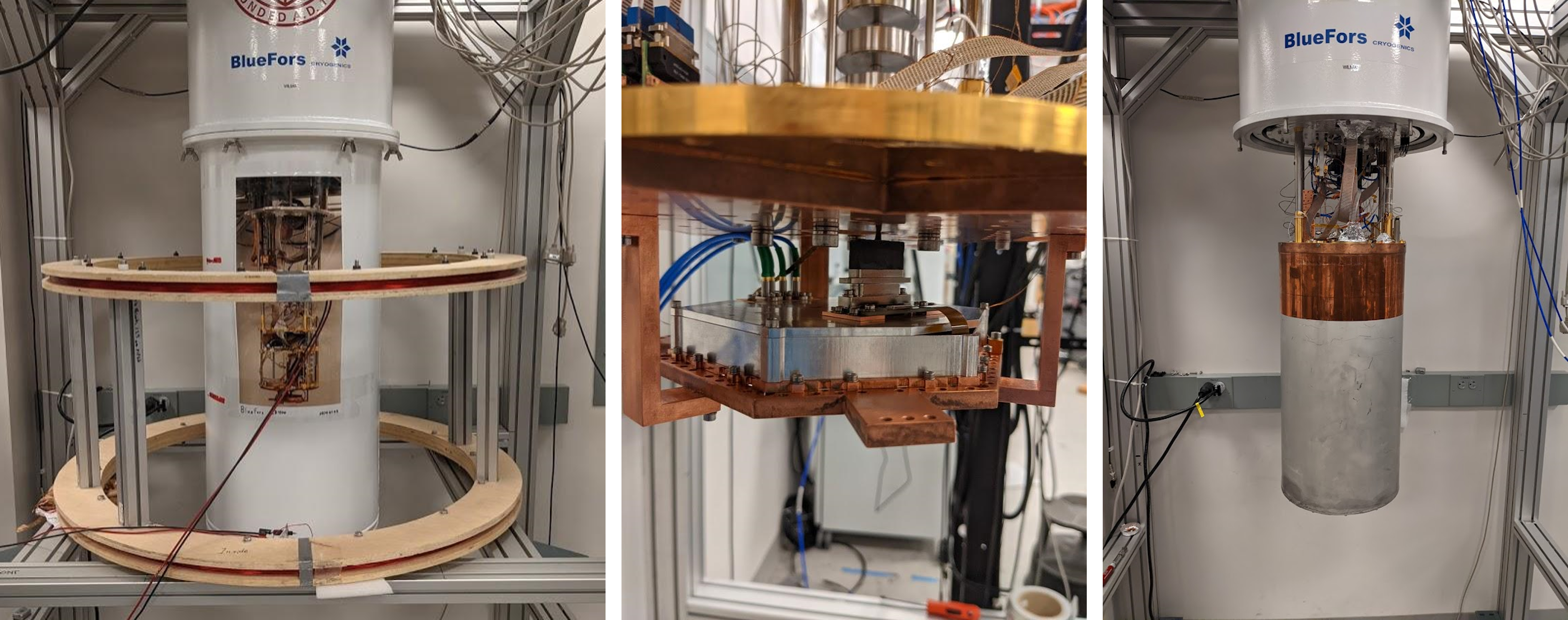}
\caption{(Color figure online.) The orientation of the Helmholtz coil around the DR (\textit{left}) and the mounting of the UMM on the bottom of the DR mixing chamber plate (\textit{center}) are aligned so the magnetic field passes perpendicularly through the plane of the UMM readout components. The tin/lead-plated 1K DR shell is also shown in position (\textit{right}) to show the extent of coverage with which it surrounds the UMM.
\vspace{-0.2in}
\label{fig:DRconfig}}
\end{center}
\end{figure}

After the readout components were cold, the external shield was removed and a set of Helmholtz coils were installed around the DR. These coils are controlled by a constant current source and provide constant DC fields perpendicular to the plane of the UMM, and thus the SQUIDs, as shown in Fig.~\ref{fig:DRconfig}. 

For each data taking run, we first applied a field of 1.6\,G for about a minute to settle the resonators, a technique found in \cite{vavagiakis20} to produce more consistent results for this system. The field was then switched to -0.26\,G without ramping to that value, data were taken, and the field was switched to the next value. For each configuration tested, data were taken at ten evenly spaced intervals from -0.26\,G to 0.26\,G. The current supply only supplies positive current, so we flipped the polarity of the connector to swap the field polarity. These field values are comparable to the $\sim$0.5 G strength of Earth's magnetic field expected on the telescope, especially given that the field will be modulated by the scanning frequency of the telescope.

In order to remove background magnetic fields and a discontinuity in measured phase shift of resonators between the positive and negative setups, we measured the phase shift at zero applied field in each configuration and subtracted it from subsequent measurements.

The data taken at each field value consisted of measuring the frequency of each SQUID channel as the flux through it was driven through four $\Phi_{0}s$ to get an f-$\Phi$ curve. Data collection was done with the SMuRF hardware and software~\cite{henderson20} that SO is using for warm readout. The data were then fit with a series of harmonics of a sine wave to extract the phase shift of the SQUID-resonator channel. As the magnetic field changes, the fractional phase shift tracks the magnetic field pickup in the channel.
We excluded from the analysis channels that were identified by the SMuRF but whose response clearly showed they were not real resonators or resonators with known fabrication issues like junction resonances (present in a small number of channels in this prototype but not in other UMMs). We also excluded channels that deviated significantly from the expected linear response to applied fields to remove channels where the SQUID acquired enough magnetic flux to wrap back around to zero phase or that do not represent the ordinary behavior of these devices.

In addition to testing different configurations of the UMM packaging, we also tested the effect of adding a superconducting shield around the UMM. This simulates the option currently being explored of adding a tin/lead-plated shell around the readout and detector electronics in both the LAT and SAT cryostats. We coated the sides and bottom of the 1K copper shell from our dilution refrigerator with a 90/10 tin/lead alloy (see Fig.~\ref{fig:DRconfig}) and repeated our pickup measurements on the aluminum clamp configuration of the UMM. For these measurements, we explored both the same range of magnetic fields as the earlier testing and larger fields up to $\pm$10.5\,G. While these field are substantially stronger than those expected from Earth's field on the telescope, we will be integrating for years while scanning through Earth's field to obtain a large signal-to-noise ratio. For the much shorter integration time in the lab, it is appropriate to apply larger fields to measure the pickup more accurately.

Since the UMM will always be coupled to the detector wafers and aluminum feedhorns during telescope operations, we also tested the magnetic pickup in a UMM once it had been coupled to detectors and feedhorns to make a UFM. For all of these measurements, the TESs were kept in the superconducting state in order to study the pickup in the SQUIDs of the UMM by themselves in a more realistic deployment configuration. We tested the UFM both with and without the tin/lead-plated shell.

\vspace{-0.2in}

\section{Results}

\vspace{-0.1in}

\subsection{Magnetic Sensitivity of UMM Configurations}

\vspace{-0.1in}

To measure the average sensitivity of a readout channel in the UMM to external magnetic fields, we averaged together the magnitude of pickup in each channel for every field value and fit the resulting data with a line for the positive applied fields and the negative applied fields separately. The slope of that fit is the amount of phase shift per unit of applied magnetic field, which is the sensitivity. 

For the aluminum configuration, the empirically measured sensitivity is 0.039 $\pm$ 0.003 $\Phi_{0}$/G for the positive fields and 0.038 $\pm$ 0.003\,$\Phi_{0}$/G for the negative fields in the first run. For the second run, the data for both the positive and negative field values yield a sensitivity of 0.054 $\pm$ 0.004\,$\Phi_{0}$/G. The positive and negative field values result in consistent fits for the second run. These results are averages across 593 channels (out of 619 channels that show real SQUID response) and 429 channels (out of 436), respectively. The two runs took place in different cooldowns with slightly different thermal properties and slightly modified RF packaging, which contributes to the different number of channels found between the two runs.
The quoted uncertainties are statistical uncertainties from fitting the sensitivities at plus and minus one standard error. They do not account for systematic uncertainties like variations in the magnetic field in our cryostat due to inhomogeneities in the Helmholtz coil field, small differences in the alignment from cooldown to cooldown, and minor repairs and modifications made to the UMM between cooldowns. These sources of systematic uncertainty likely account for the difference in sensitivity between the two runs.

\begin{figure}[htbp]
\begin{center}
\includegraphics[width=1.0\linewidth, keepaspectratio]{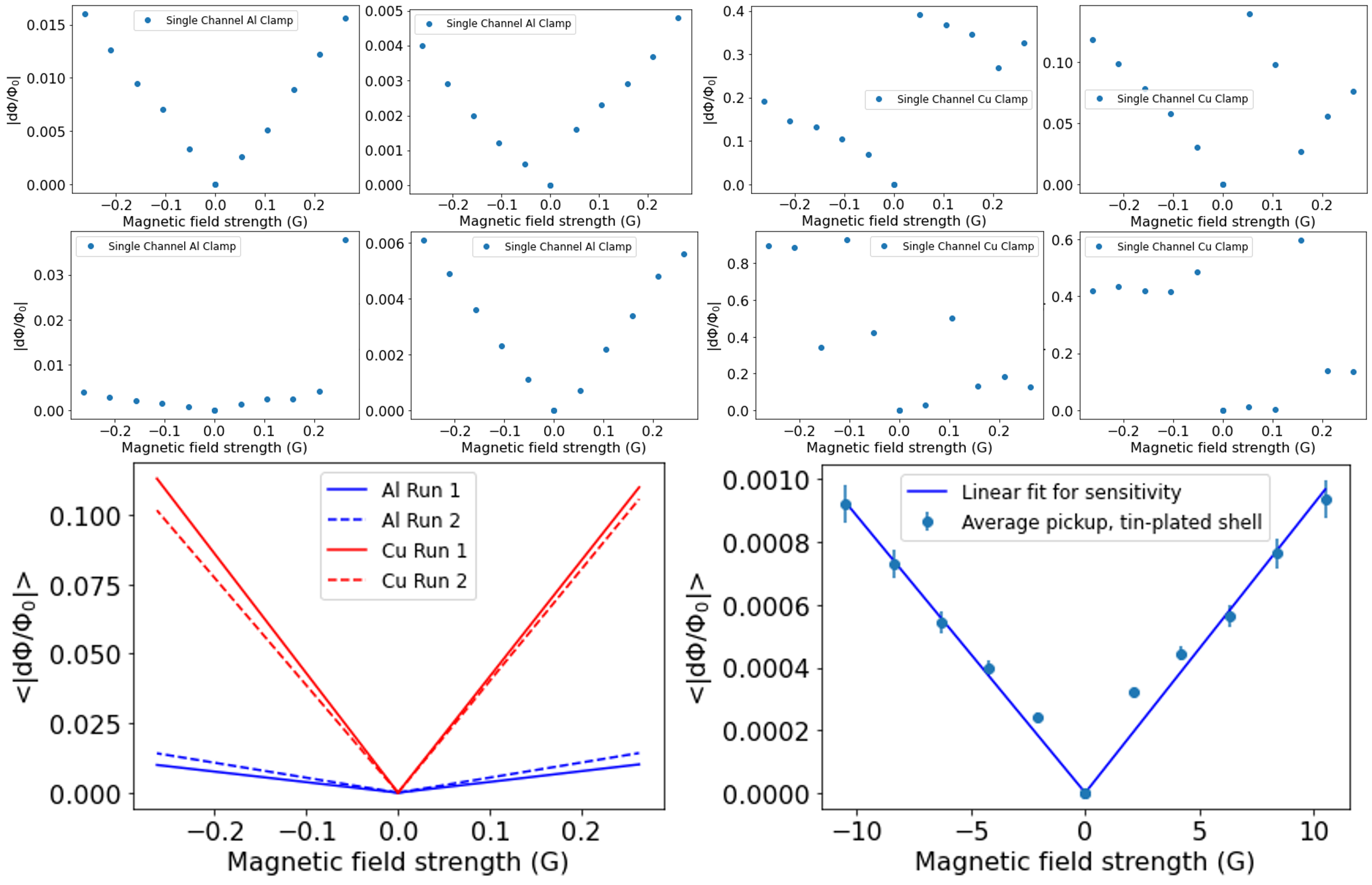}
\caption{(Color figure online.) \textit{Top:} Magnetic pickup given as the fractional phase shift for four resonators for a test of the Al clamp (\textit{top left}) and the same resonators in a test of the Cu clamp (\textit{top right}) during cooldowns with similar overall noise properties. With the Cu clamp, the resonators experience a significantly higher magnitude of pickup which varies from resonator to resonator and deviations from a linear sensitivity at different applied field values. While such jumps can occur in the Al clamp configuration (as in the bottom left plot), they are comparatively rare. We hypothesize that this higher pickup with the Cu clamp is due to a greater sensitivity to the lab environment. \textit{Bottom Left:} Average phase shift of those channels on the UMM with linear response as a function of magnetic field. Two runs are shown for the copper clamp ({\it red, solid and dashed}) and two for the aluminum clamp ({\it blue, solid and dashed}). The aluminum clamp clearly has less magnetic field pickup for these channels. \textit{Bottom Right:} Data with standard errors and fits for the Al clamp with the tin/lead-plated shell installed. Even at these significantly larger field values, the pickup remains small.  
\vspace{-0.2in}
\label{fig:channels}}
\end{center}
\end{figure}

Measurements of the copper configuration were more complex to interpret than the aluminum configuration. The overall shielding environment was less stable than the aluminum configuration. The SQUID curves themselves were noisier, and many channels often appeared to include random pickup (either an excess or a deficit) for only a few of the field values while the remaining values followed a somewhat linear trend (see Fig.~\ref{fig:channels}). We believe that this behavior is related to decreased RF stability of the device and an increased sensitivity to RF and magnetic signals from outside the cryostat other than the applied magnetic field. There was also a significant increase in the number of channels with high enough magnetic field pickup to undergo wrapping in $\Phi_{0}$ for one or more of the magnetic field values at which data was taken. Taken together, these differences mean that far fewer channels passed when applying the same cuts as we did to the aluminum configuration.

Of the remaining channels, the measured sensitivity is considerably higher than for the aluminum configuration. For the first copper test, the average sensitivity across 70 channels (out of 459) was 0.42 $\pm$ 0.03\,$\Phi_{0}$/G for positive field values and 0.43 $\pm$ 0.04\,$\Phi_{0}$/G for negative field values. A second cooldown in which the UMM exhibited better overall noise performance yielded 0.40 $\pm$ 0.03\,$\Phi_{0}$/G for positive field values and 0.39 $\pm$ 0.04\,$\Phi_{0}$/G for negative field values across 61 channels (out of 562). Fig.~\ref{fig:channels} (bottom left) plots the sensitivities for the copper and aluminum clamps on the same axis for comparison.

The cuts for linearity may remove some of the channels that exhibit high sensitivity to magnetic fields. Even so, since the largest number of cuts occurred for the copper clamp, we expect that any reduction in the number of cut channels would only increase the advantage seen in the shielding factor for the aluminum clamp compared to the copper clamp since the measured sensitivity with the copper clamp sensitivity would increase more than the sensitivity measured with the aluminum clamp.

\vspace{-0.2in}

\subsection{Tin/Lead-plated 1K Shield}
\label{sec:tin-lead}
\vspace{-0.1in}

We repeated our pickup measurements in the aluminum UMM configuration with a 1K DR shell plated with a 90/10 tin/lead alloy installed (Fig.~\ref{fig:DRconfig}). Initial measurements over the same range of magnetic field values as before (-0.26 to 0.26\,G) showed erratic pickup nearly two orders of magnitude smaller than without the tin/lead-plated shell. After using a larger range of field values (-10.5\,G to 10.5\,G), a linear response was recovered with sensitivity $9.2 \times 10^{-5}$ $\pm$ $0.6 \times 10^{-5}$\,$\Phi_{0}$/G for positive field values and $8.8 \times 10^{-5}$ $\pm$ $0.7 \times 10^{-5}$\,$\Phi_{0}$/G for negative field values with 659 channels (out of 683) (Fig.~\ref{fig:channels}, bottom right).

\vspace{-0.2in}

\subsection{Magnetic Sensitivity of the UMM with Feedhorns Installed}

\vspace{-0.1in}

\begin{figure}[htbp]
\begin{center}
\includegraphics[width=1.0\linewidth, keepaspectratio]{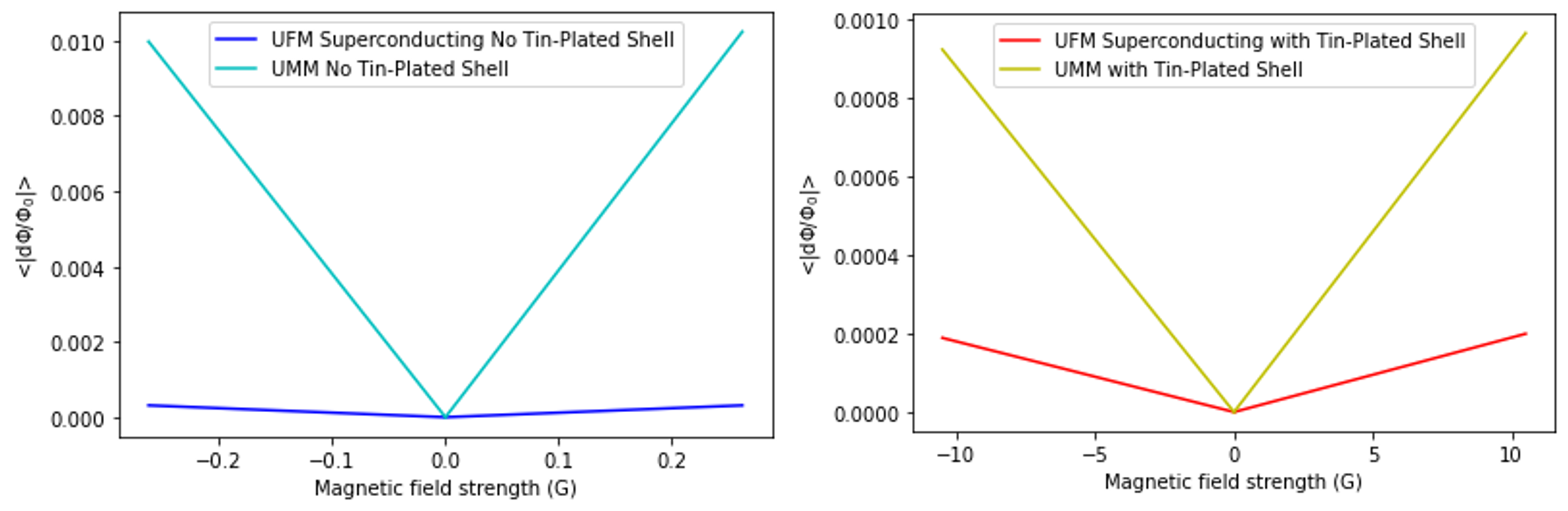}
\caption{(Color figure online.) \textit{Left:} Magnetic pickup given as the fractional phase shift for a UMM (cyan) and for a UFM with all TESs superconducting (blue) without the tin/lead-plated shell installed. The addition of the feedhorns to the UMM leads to a factor of $\sim$30 improvement in pickup. \textit{Right:} Magnetic pickup given as the fractional phase shift for a UMM (yellow) and for a UFM with all TESs superconducting (red) with the tin/lead-plated shell installed. The addition of the feedhorns to the UMM leads to a factor of $\sim$5 improvement in pickup.  
\vspace{-0.2in}
\label{fig:UMMvUFM}}
\end{center}
\end{figure}

Using a UFM with a gold-plated aluminum clamp and aluminum feedhorns, we tested the magnetic sensitivity of the SQUIDs in the UMM with the detectors unbiased to see how the feedhorns would affect the magnetic shielding. We took three sets of data with the UFM in the superconducting state without the tin/lead-plated shell installed to test the repeatability of our measurements. For the first run, both halves of the UFM were operational. Across 1277 channels, the sensitivity for the positive field values was $1.05 \times 10^{-3}$ $\pm$ $0.09 \times 10^{-3}$\,$\Phi_{0}$/G and the sensitivity for the negative field values was $1.02 \times 10^{-3}$ $\pm$ $0.09 \times 10^{-3}$\,$\Phi_{0}$/G. For the other two runs in a subsequent cooldown, only one half of the UFM was operable due to issues with the RF connection on the UFM. For the first of these runs, the sensitivity for both the positive and negative field values was $1.4 \times 10^{-3}$ $\pm$ $0.2 \times 10^{-3}$\,$\Phi_{0}$/G for the 474 channels passing cuts. The second set of data from later in that same cooldown but with a factor of 10 stronger magnetic fields yielded a sensitivity for the positive field values of $1.0 \times 10^{-3}$ $\pm$ $0.1 \times 10^{-3}$\,$\Phi_{0}$/G and for the negative field values of $1.1 \times 10^{-3}$ $\pm$ $0.1 \times 10^{-3}$\,$\Phi_{0}$/G for 469 channels passing cuts. Averaging these results together gives a sensitivity of $1.2 \times 10^{-3}$ $\pm$ $0.2 \times 10^{-3}$\,$\Phi_{0}$/G for both the positive and negative field values, which represents a $\sim$30 factor of improvement over the pickup for the UMM by itself (Fig.~\ref{fig:UMMvUFM}, left).

We also tested the pickup in the UFM with detectors unbiased while the tin/lead-plated shell was installed. Using the same range of field values as in Section~\ref{sec:tin-lead}, we measured a sensitivity of $1.9 \times 10^{-5}$ $\pm$ $0.2 \times 10^{-5}$\,$\Phi_{0}$/G for the positive field values and $1.8 \times 10^{-5}$ $\pm$ $0.2 \times 10^{-5}$\,$\Phi_{0}$/G for the negative field values across 537 channels passing cuts. This represents a factor of $\sim$5 improvement over the UMM without feedhorns (Fig.~\ref{fig:UMMvUFM}, right).

\vspace{-0.2in}

\subsection{Spatial Variation of Pickup}

\vspace{-0.1in}

\begin{figure}[htpb]
\vspace{-0.2in}
\centering
\includegraphics[width=0.9\linewidth]{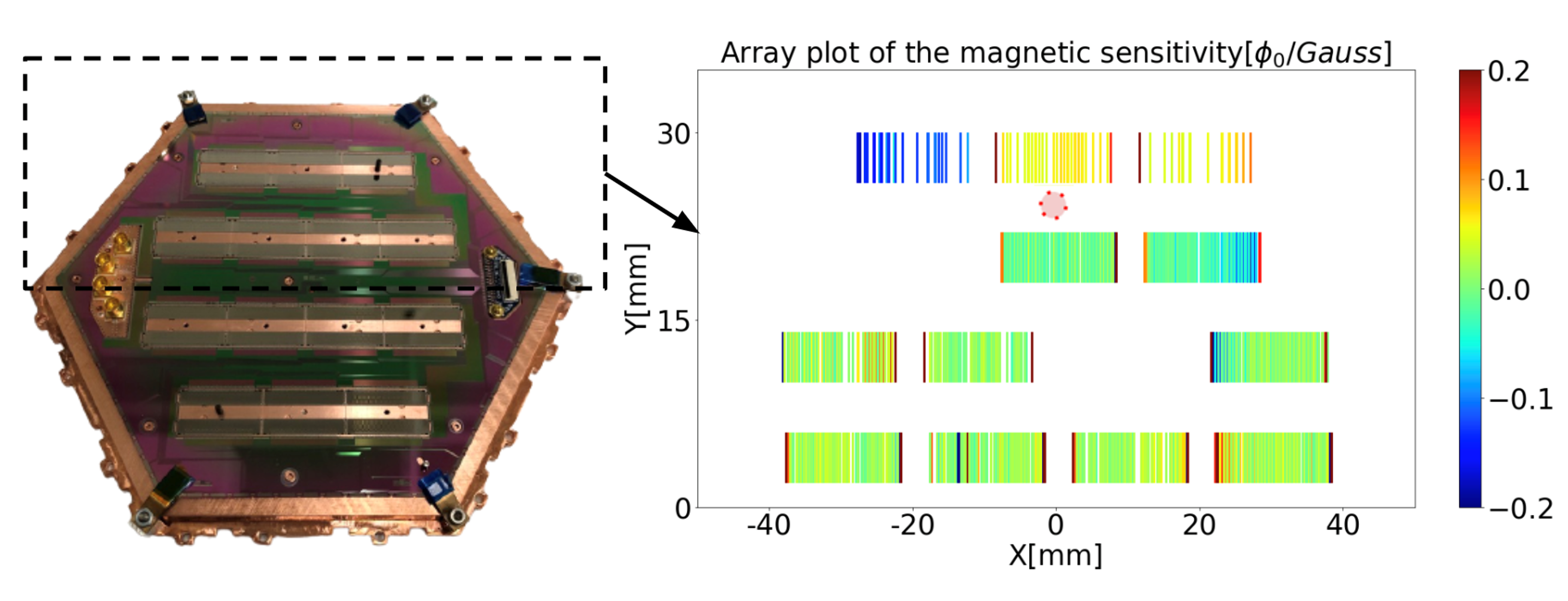}
\caption{(Color figure online.) {\it Left}: Photograph of the UMM \cite{ehealy20} without the outer clamp and copper lid. The UMM used for
these measurements was missing two of the 14 $\mu$mux chips (shown in green in the picture directly above and below the strips of copper in the middle of the UMM) that a UMM would usually have when tested. {\it Right}: Plot of the magnetic sensitivity of each SQUID channel (shown as a vertical line based on location on the mux chip) in\,$\Phi_{0}$/G. The red hexagon indicates the location of the screw that was changed for the measurements described in the text.
\vspace{-0.2in}
\label{fig:array_plot}}
\end{figure}

Apart from the overall sensitivity with different magnetic shielding configurations, we also studied the spatial variation of pickup from individual resonator channels across the array. Fig.~\ref{fig:array_plot} shows the array plot of the magnetic sensitivity of each resonator channel when the prototype UMM is assembled with the aluminum clamp. Higher pickup is found on the resonator channels near the outside of the UMM and the edges of the $\mu$mux chips. The latter is due to the asymmetry of the environment near edge channels as discussed in \cite{connors21}.

While all the screws inside the footprint of the UMM are brass, stainless steel type 304 screws are used for the exterior vertices. To test if these screws were causing the higher pickup around the outside of the UMM, we compared the magnetic sensitivity before and after switching one of the internal brass screws to a stainless steel one with the expectation that this would increase the pickup near that screw. No significant change in pickup was observed in channels physically close to the screw, so it is unlikely that the hardware is causing significant magnetic pickup. It is possible that the higher pickup around the edges reflects the complicated shape of the magnetic fields as they are routed around the superconducting clamp.

\vspace{-0.2in}

\section{Conclusion}

\vspace{-0.1in}

We measured magnetic pickup in a prototype SO UMM with different packaging, or clamps, as well as inside a 90/10 tin/lead plated shell. The aluminum clamp provides both a more stable magnetic and RF environment and an improvement in shielding factor of eight to ten relative to the copper clamp. It is clear that the aluminum clamp outperforms the copper clamp, thereby confirming that the choice of aluminum for the outer UMM clamp is preferred over copper.

The spatial distribution of the pickup is consistent with the UMM clamp and the $\mu$mux chip ground plane providing better shielding near their geometric centers with a modest degradation in shielding near the edges.

These results also show that adding a 90/10 tin/lead plated shield at 1\,K around the readout and detectors can significantly reduce the overall level of magnetic pickup. We measure a shielding factor improvement $>$100
relative to the aluminum clamp with no tin/lead-plated shell. Such a shell for the telescopes would need to have optical access to the detectors and more feedthroughs for readout electronics, so the deployed shielding factor may be degraded compared to these measurements. 
Based on these results, the same ratio of 90/10 tin/lead plating will be applied to 1\,K copper shells surrounding the detectors and readout in both the SO SATs and the LAT receiver.

Tests of an SO Universal Focal Plane Module (UFM) that includes detectors and aluminum feedhorns along with the UMM readout while the detectors are in the superconducting state show a further improvement in shielding of $\sim$30 without the tin/lead-plated shell installed and of $\sim$5 with the tin/lead-plated shell installed compared to the UMM alone, suggesting that the feedhorns reduce the amount of magnetic pickup in the SQUIDs of the UMM. Future studies of the magnetic pickup of the UFM might include characterization of TES pickup levels while biased on the transition and in the normal state to further explore the effects of magnetic pickup in the detectors while operating at the telescope.

\begin{acknowledgements}
This work was supported in part by a grant from the Simons Foundation (Award \#457687, B.K.). ZBH acknowledges support from a NASA Space Technology Graduate Research Opportunities Award. SKC acknowledges support from NSF award AST-2001866. ZX is supported by the Gordon and Betty Moore Foundation through grant GBMF5215 to the Massachusetts Institute of Technology.
\end{acknowledgements}

\pagebreak

\end{document}